\documentstyle[11pt]{article}
\def\fnote#1#2{\begingroup\def\thefootnote{#1}\footnote{#2}\addtocounter
{footnote}{-1}\endgroup}

\begin{document}

\hfill{UTTG-08-08}

\vspace{36pt}

\begin{center}
{\large {\bf {Non-Gaussian Correlations Outside the Horizon  II: The General Case}}}

\vspace{36pt}
Steven Weinberg\fnote{*}{Electronic address:
weinberg@physics.utexas.edu}\\
{\em Theory Group, Department of Physics, University of
Texas\\
Austin, TX, 78712}

\vspace{30pt}

\noindent
{\bf Abstract}
\end{center}
\noindent
The results of a recent paper [0808.2909] are generalized.  A more detailed proof is presented that under essentially all conditions, the non-linear classical equations governing matter and gravitation in cosmology have ``adiabatic'' solutions in which, far outside the horizon, in a suitable gauge, the reduced spatial metric $g_{ij}({\bf x},t)/a^2(t)$ becomes a 
time-independent function ${\cal G}_{ij}({\bf x})$, and all perturbations to the other metric components and to all matter variables vanish.  The corrections are of order $a^{-2}$, and their ${\bf x}$-dependence is now explicitly given in terms of ${\cal G}_{ij}({\bf x})$ and its derivatives.  The previous results for the time-dependence of the corrections to $g_{ij}({\bf x},t)/a^2(t)$ in the case of multi-scalar field theories are now shown to apply for any theory whose anisotropic inertia vanishes to order $a^{-2}$.  
Further, it is shown that the adiabatic solutions are attractive as $a$ becomes large for the case of single field inflation and now also for thermal equilibrium with no non-zero conserved quantities, and the $O(a^{-2})$ corrections to the other dynamical variables are explicitly calculated in both cases.  

 \vfill

\pagebreak

\begin{center}
{\bf I. Introduction}
\end{center}

In a recent paper [1], I addressed the problem of the evolution of non-Gaussian correlations during the long period when the observationally interesting wave lengths were outside the horizon, and the universe  passed through transitions such as  reheating and baryonsynthesis that are not well understood.  It was shown on general grounds that, apart from quantum corrections, in a suitable coordinate system, the exact non-linear field equations always have an adiabatic solution for which  $g_{ij}/a^2$ becomes time independent outside the horizon, and the  perturbations to $g_{00}$ and $g_{i0}$ and to all matter variables become negligible.  This, however, is only part of the problem.  In order to conclude in some class of  theories that cosmological  perturbations  were really described by the adiabatic solution throughout the time when  they  were outside the horizon, it is necessary to show that for these theories the adiabatic solution is attractive in the limit of large $a(t)$.  In [1] the corrections to this solution for the metric were calculated explicitly for a theory with any number of scalar fields and an arbitrary potential, and it was shown that for all such theories this solution is attractive as far as the spatial part of the metric is  concerned, but this could be shown for    the scalar field perturbations only in  the case of single field inflation [2].   The present paper clarifies these general arguments in Section II, and shows in Section III that the explicit solution for the metric obtained in [1] applies to any theory that (like scalar field theories) has vanishing anisotropic inertia to lowest order in perturbations.  Sections IV--VI present an an improved argument  that this solution is attractive for all metric and matter variables,   not only in the case of single field inflation, but also for a sufficiently long period of local thermal equilibrium with no non-zero conserved quantities.

\begin{center}
{\bf II. The General Adiabatic Solution}
\end{center}

In this section we will give a general broken-symmetry argument that shows the existence of certain adiabatic solutions of the non-linear field equations for the metric and matter variables.  The discussion will follow along the same lines as in Section II of [1],  but we will here obtain more detailed results, that are of some interest in themselves, and will be used in the next section.

Before proceeding to this task, it should be emphasized that the universal existence of adiabatic solutions of the field equations does not in itself mean that these are the solutions that apply to the fluctuations in matter and gravitation actually present in our universe.  We will be able to show in Sections IV--VI that these solutions are attractive in two cases: single field inflation, and local thermal equilibrium with no non-zero conserved quantities.  Based on our experience with the linearized equations, it is plausible that fluctuations in these cases are in the basin of attraction of the adiabatic solutions, but we will not try to prove this for the non-linear equations.

Turning to the existence of the adiabatic solutions, our discussion in this section is based on two very general assumptions:
\begin{enumerate}
\item We assume that the dynamical equations governing matter and gravitation are generally covariant.  This of course is just a way of implementing the Principle of Equivalence.
\item 
We assume that whatever the dynamical equations governing the metric and matter (including radiation) variables may be, these equations have a solution in which the metric takes the flat-space Robertson--Walker form, with $g_{00}=-1$, $g_{0i}=0$, and $\tilde{g}_{ij}\equiv g_{ij}/a^2 =\delta_{ij}$, and in which all matter variables take their unperturbed form; that is, all densities and pressures and scalar fields are functions only of time, and all velocities and other 3-vectors vanish.
\end{enumerate}  

We will show that under these conditions,  for a suitable choice of spacetime coordinates, these equations  always also have a family of solutions that  for large $a(t)$ have the following properties:
\begin{enumerate}
\item The metric for any of these solutions has components with 
\begin{eqnarray}
&&g_{ij}({\bf x},t)= a^2(t)\Big[{\cal G}_{ij}({\bf x})+f_1(t){\cal R}_{ij}({\bf x})+f_2(t){\cal G}_{ij}({\bf x}){\cal R}({\bf x})+\dots \Big]\nonumber\\&& g_{i0}({\bf x},t)=f_3(t)\partial_i{\cal R}({\bf x})+\dots \;,\\&&
g_{00}({\bf x},t)=-1+f_4(t){\cal R}({\bf x})+\dots \;,\nonumber
\end{eqnarray}
\item Whether or not the energy and momentum of any particular constituent of the universe is separately conserved, for this solution its energy-momentum tensor has the form
\begin{eqnarray}
&&T_{ij}({\bf x},t)=a^2(t)\Big[{\cal G}_{ij}({\bf x})\bar{p}(t)+g_1(t){\cal R}_{ij}({\bf x})+g_2(t){\cal G}_{ij}({\bf x}){\cal R}({\bf x})+ \dots\Big]\;,\nonumber\\&&T_{i0}({\bf x},t)=g_3(t)\partial_i{\cal R}({\bf x})+\dots\;,\\&&T_{00}({\bf x},t)=\bar{\rho}(t)+g_4(t){\cal R}({\bf x})+\dots\;,\nonumber
\end{eqnarray}
\item
Four-scalars  $s({\bf x},t)$ such as  temperatures, number densities, or scalar fields, have the form
\begin{equation}
s({\bf x},t)=\bar{s}(t)+h(t){\cal R}({\bf x})+\dots\;.
\end{equation}
\end{enumerate}
Here ${\cal G}_{ij}({\bf x})$ is an arbitrary positive matrix function only of the spatial coordinates, which can conveniently be taken as the value of $g_{ij}({\bf x},t)/a^2(t)$ at some time $t=
T$; ${\cal R}_{ij}({\bf x})$ and ${\cal R}({\bf x})$ are  the Ricci tensor and curvature scalar for the metric ${\cal G}_{ij}({\bf x})$;  the functions $f_n(t)$, $g_n(t)$ and $h(t)$  are of order $a^{-2}(t)$;  dots denote terms of higher order in $1/a(t)$; and a bar over any quantity indicates its unperturbed value.    Although these solutions exist for any ${\cal G}_{ij}({\bf x})$, quantum fluctuations in inflation produce a particular stochastic ${\cal G}_{ij}({\bf x})$, whose properties will not concern us here.  Aside from quantum loop effects, a term of order $a^{-n}$ makes a contribution to correlation functions with characteristic wave number $k$ that is expected to be suppressed outside the horizon by factors of order $(k/aH)^n$, where as usual $H\equiv \dot{a}/a$.

To illustrate what is meant by functions of order $a^{-2}(t)$, we note here that, as we will show in the next section, in suitable coordinate system, as long as there is no anisotropic inertia to order $a^{-2}$, the functions $f_1(t)$ and $f_2(t)$ are in general given by
\begin{eqnarray}
&&f_1(t)=2\int^t_T\frac{dt'}{a^3(t')}\int_{T}^{t'} a(t'')\,dt''\\
&&f_2(t)=-\frac{1}{2}\int^t_T\frac{dt'}{a^3(t')}\int_{T}^{t'} a(t'')\,dt''+\frac{1}{2}\int_T^t\frac{H^2(t')\,dt'}{\dot{H}(t')a^3(t')}\int_{T}^{t'} a(t'')\,dt''\nonumber\\
&&~~~~~-\frac{1}{2}\int_T^t \frac{H(t')\,dt'}{a^2(t')\,\dot{H}(t')}\;,
\end{eqnarray}
while the coordinate system is defined so that $f_3(t)=0$.  Here $T$ is an arbitrary time, at which we choose $g_{ij}/a^2$ to equal ${\cal G}_{ij}$.
For instance, during the radiation-dominated era when $a\propto t^{1/2}$, at a time $t\gg T$, these give
\begin{eqnarray}
&&f_1(t)\rightarrow\frac{1}{3a^2(t)H^2(t)}\\
&&f_2(t)\rightarrow 0\;.
\end{eqnarray}
The space derivatives in ${\cal R}_{ij}({\bf x})$ contribute two factors of a characteristic wave number $k$, so that these $O(a^{-2})$ terms do indeed make contributions of order $(k/aH)^2$.

On the other hand, the function $f_4(t)$ depends on the nature of the constituents of the universe.  We show in Section V that during single field inflation
\begin{equation}
f_4(t)=-\frac{1}{2\dot{H}(t)}\left(-\frac{1}{a^2(t)}+\frac{H(t)}{a^3(t)}\int_T^t a(t')\;dt'\right)\;,
\end{equation}
while in Section VI we show that in a period of thermal equilibrium with no non-zero conserved quantities,
\begin{eqnarray}
&&f_4(t)=\frac{\dot{H}(t)}{H^2(t)}\int_T^t \left(\frac{H(t')}{\dot{H}(t')}+\frac{\ddot{H}(t')}{6\dot{H}^2(t')}\right)\left(-\frac{1}{a^2(t')}+\frac{H(t')}{a^3(t')}\int_T^{t'}a(t'')\,dt''\right)\;dt'\nonumber\\&&~~~~
-\frac{1}{2\dot{H}(t)}\left(-\frac{1}{a^2(t)}+\frac{H(t)}{a^3(t)}\int_T^{t}a(t')\,dt'\right)\;.
\end{eqnarray}

The first step in proving the existence of these solutions is to consider instead a trial configuration, in which $g_{ij}({\bf x},t)=a^2(t){\cal G}_{ij}({\bf x})$, where ${\cal G}_{ij}({\bf x})$ is an arbitrary positive matrix function of position, and all other variables are the same as for the Robertson--Walker solution; that is, $g_{00}=-1$, $g_{i0}=0$; all densities and pressures take their unperturbed values; and all velocities vanish.  Note that this is not a small perturbation to the Robertson--Walker solution, for no assumption will be made that ${\cal G}_{ij}({\bf x})$ is close to $\delta_{ij}$.  If ${\cal G}_{ij}({\bf x})$ were constant then this would be an exact solution, since it could be obtained from the Robertson--Walker solution by a coordinate transformation $x^i\rightarrow A^i{}_jx^j$, where $A^k{}_iA^k{}_j={\cal G}_{ij}$.  Thus we expect the trial configuration to fail to satisfy the field equations only by terms involving derivatives of ${\cal G}_{ij}({\bf x})$.  These terms must be accompanied with factors of $1/a$, as required by the condition that these field equations must be invariant under the scale transformations
\begin{equation}
 x^i\rightarrow \lambda x^i\;,~~~~~~~~a(t)\rightarrow a(t)/\lambda\;,
\end{equation}
under which $g_{ij}/a^2$, $g_{i0}/a$, $g_{00}$, $T_{ij}/a^2$, $T_{i0}/a$, $T_{00}$ (for each constituent of the universe), 
and all pressures, densities, temperature, etc. are invariant.  (This is just a different way of expressing invariance under the coordinate transformations   $x^i\rightarrow \lambda x^i$, under which $g_{\mu\nu}$ and $T_{\mu\nu}$ transform as tensors, and $a(t)$ is invariant.)  Hence for sufficiently large $a(t)$, it is a reasonable ansatz to seek a solution in which deviations of all scale-invariant metric and matter variables from the trial configuration are small perturbations.

 The field equations for the complete set of scale-invariant perturbations then take a form that can be symbolized as $Dq=S$, where $q$ is a column of scale-invariant perturbations, $D$ is a matrix formed from a linear combination of first and higher time derivatives whose coefficients are scale-invariant ${\bf x}$-independent functions of time, and $S$ is a column of  source terms, given by a linear combination of appropriate ${\bf x}$-dependent 3-tensors formed from derivatives of ${\cal G}_{ij}({\bf x})$, with factors of $a(t)$ as required to make the source terms scale-invariant.    
Concrete examples will be given in Sections III, V, and VI.  But even without going into the details of these equations, we can see the general features of a class of  solutions.  If we make the ansatz that the solution for each perturbation is a linear combination of the same ${\bf x}$-dependent 3-tensors that appear in the source terms for this perturbation, with coefficients given by functions only of time, then the dynamical equations reduce to a set of coupled ordinary inhomogeneous differential equations for these coefficient functions, with source terms having whatever powers of $1/a(t)$ were required by scale invariance in the original equations.  Such equations will always have an adiabatic  solution, in which the functions of time accompanying each 3-tensor are of an order in $1/a(t)$ that is the same as the number of powers of  $1/a(t)$ in the corresponding source term, plus some number of solutions of the homogeneous equations $Dq=0$ that may in general increase or decrease  as $a(t)$ increases.  It is the time-dependence of the solutions of the homogeneous equation that tells us whether the adiabatic solution is attractive or not, the question addressed in Sections IV through VI.

For $g_{ij}/a^2$ the source term must be a linear combination of the scale-invariant tensors $a^{-2}(t){\cal R}_{ij}({\bf x})$ and $a^{-2}(t){\cal G}_{ij}({\bf x}){\cal R}({\bf x})$, plus terms with more than two derivatives that are suppressed by more than two powers of $1/a(t)$.  For $g_{i0}/a$ the source term must be a term proportional to  the scale-invariant 3-vector $a^{-3}(t)\partial_i{\cal R}({\bf x})$, plus terms with more than three derivatives that are suppressed by more than three powers of $1/a(t)$.  For $g_{00}$ the source term must be a term proportional to  the scale-invariant scalar $a^{-2}(t){\cal R}({\bf x})$, plus terms with more than two derivatives that are suppressed by more than two powers of $1/a(t)$.  Thus we expect these equations to have solutions of the form (1).  The same reasoning applies to whatever matter variables (densities, pressures, velocities, and if necessary anisotropic inertia) that enter in the energy-momentum tensors for each constituent of the universe, giving energy-momentum tensors of the form (2) and scalars of the form (3).

It should be noted that Eq.~(2) tells us that $\delta T_{i0}$ is a gradient to order $a^{-2}$, so that to this order there is no vorticity.  On the other hand, Eq.~(2) allows terms in $\delta T_{ij}$ that are not proportional to $\delta_{ij}$, so there is no general argument against the presence of anisotropic inertia to order $a^{-2}$.  The results of [1] show that there is no anisotropic inertia to this order in  a theory of multiple scalar fields with  an arbitrary potential, but solution of the collisionless Boltzmann equation [3] shows that anisotropic inertia of  order $a^{-2}$ can be produced in the linear approximation by gravitational perturbations of this order in a gas of non-interacting relativistic particles.

\begin{center} 
{\bf III. Explicit Solution for the Metric}
\end{center}

We will now verify by detailed calculation that, under the assumptions of the previous section, for large $a(t)$ there is indeed a solution of the gravitational field equations for which, in a suitable spacetime coordinate system, the metric takes the form (1), and we will calculate the functions $f_n(t)$ appearing in this solution.
For the metric, we use the ADM parameterization [4]:
\begin{eqnarray}
&& g_{00}=-N^2+g_{ij}N^iN^j\;,~~~~g_{0i}=g_{ij}N^j\equiv N_i\nonumber\\
&& g^{00}=-N^{-2}\;,~~~~g^{i0}=N^i/N^2\;,~~~~g^{ij}={}^{(3)}g^{ij}-N^iN^j/N^2\;,
\end{eqnarray}
where ${}^{(3)}g^{ij}$ is the reciprocal of the $3\times 3$-matrix $g_{ij}$.  
It will be convenient also to write 
\begin{equation}
g_{ij}({\bf x},t)=a^2(t)\tilde{g}_{ij}({\bf x},t)\;,
\end{equation}
where $a(t)$ is the Robertson--Walker scale factor appearing in the unperturbed solution.
In this notation, and in units with $8\pi G\equiv 1$, the gravitational field equations are
\begin{equation}
\tilde{\nabla}_i\Big(C^i{}_j-\delta^i{}_jC^k{}_k\Big)=-NT^0{}_j\;,
\end{equation}
\begin{equation}
-a^{-2}\tilde{g}^{ij}\tilde{R}_{ij}-C^i{}_jC^j{}_i+(C^i{}_i)^2=2N^2T^{00}\;,
\end{equation}
\begin{eqnarray}
&&\tilde{R}_{ij}-C^k{}_kC_{ij}+2C_{ik}C^k{}_j+N^{-1}\Big(-\dot{C}_{ij}+C^k{}_i\tilde{\nabla}_jN_k+C^k{}_j\tilde{\nabla}_iN_k\nonumber\\
&&~~~+N^k\tilde{\nabla}_kC_{ij}+\tilde{\nabla}_i\tilde{\nabla}_jN\Big)=-T_{ij}+\frac{1}{2}a^2\tilde{g}_{ij}T^\lambda{}_\lambda\;,
\end{eqnarray}
where $\tilde{g}^{ij}({\bf x},t)$ is the reciprocal of the matrix $\tilde{g}_{ij}({\bf x},t)$;  $\tilde{R}_{ij}({\bf x},t)$ is the three-dimensional Ricci tensor  for the metric $\tilde{g}_{ij}({\bf x},t)$; and $C^i{}_j({\bf x},t)$ is the extrinsic curvature of the surfaces of fixed time
\begin{equation}
C^i{}_j\equiv a^{-2}\tilde{g}^{ik}\,C_{kj}\;,~~~~C_{ki}\equiv \frac{1}{2N}\Big[2a\dot{a}\tilde{g}_{ki}+a^2\dot{\tilde{g}}_{ki}-\tilde{\nabla}_k N_i-\tilde{\nabla}_i N_k\Big]\;,
\end{equation}
where $\tilde{\nabla}_i$ is the three-dimensional covariant derivative calculated with the three-metric $\tilde{g}_{ij}$.  

In accordance with the remarks of the previous section, we tentatively assume a solution in which $\dot{\tilde{g}}_{ij}$ and $g_{00}+1$ are small perturbations for large $a(t)$, of order $1/a^2(t)$.  We also adopt a definition of space coordinates for which $N^i=0$, so that $g_{i0}=0$.  We  then have 
\begin{equation}
C^i{}_j=H\delta^i{}_j+\xi^i{}_j\;,
\end{equation}
where $\xi^i{}_j$ is, like $\dot{\tilde{g}}_{ij}$ and $\delta N$, a quantity whose leading term is of order $a^{-2}$:
\begin{equation}
\xi^i{}_j=\frac{1}{2}\tilde{g}^{ik}\left[\dot{\tilde{g}}_{kj}-2H\,\delta N\,\tilde{g}_{ki}\right]+O(a^{-4})\;.
\end{equation}
Also in accordance with what we found in the previous section, the total energy momentum tensor for this solution takes the form
\begin{equation}
T_{ij}=a^2\bar{p}\tilde{g}_{ij}+\delta T_{ij}\;,~~~~T_{i0}=\delta T_{i0}\;~~~~T_{00}=\bar{\rho}+\delta T_{00}\;,
\end{equation}
where  $\bar{\rho}=3H^2$, $\bar{p}=-2\dot{H}-3H^2$,  and  $\delta T_{ij}/a^2$, $\delta T_{i0}$, and $\delta T_{00}$ are all of order $a^{-2}$.  
The gravitational field equations now read
\begin{equation}
\tilde{\nabla}_i\Big(\xi^i{}_j-\delta^i{}_j\xi^k{}_k\Big)=\delta T_{0j}+O(a^{-4})\;,
\end{equation}
\begin{equation}
-12H^2\delta N+2\delta T_{00}=-a^{-2}\tilde{g}^{ij}\tilde{R}_{ij}+4H\xi^k{}_k+O(a^{-4})\;,
\end{equation}
\begin{eqnarray}
&&\dot{\xi}^i{}_j+3H\xi^i{}_j+H\delta^i{}_j\xi^k{}_k+(3H^2-\dot{H})\delta  N\delta^i{}_j=a^{-2}\tilde{g}^{ik}\tilde{R}_{kj}
\nonumber\\&&
~~~~~+a^{-2}\tilde{g}^{ik}\delta T_{kj}-\frac{1}{2}a^{-2}\delta^i{}_j\tilde{g}^{kl}\delta T_{kl}+\frac{1}{2}\delta^i{}_j\delta T_{00}
+O(a^{-4})\;.
\end{eqnarray}

Now, as remarked in Section II, to order $a^{-2}$ these solutions have no vorticity, so we can introduce a momentum potential $U$, of order $a^{-2}$,
such that 
\begin{equation}
\delta T_{i0}=\partial_i U\;.
\end{equation}
The equation of momentum conservation then reads
\begin{equation}
a^{-2}\tilde{g}^{ik}\tilde{\nabla}_i\delta T_{kj}=2\dot{H}\partial_j\delta N+\partial_j(\dot{U}+3HU)+O(a^{-4})\;.
\end{equation}
We can always also  write
\begin{equation}
\delta T_{ij}=a^2\Big(\tilde{g}_{ij}\delta p+\Pi_{ij}\Big)\;,
\end{equation}
where $\Pi_{ij}$ is a 3-tensor of order $a^{-2}$ representing anisotropic inertia, with $\tilde{g}^{ij}\Pi_{ij}=0$.  Eq.~(2) for the total energy momentum tensor together with the Bianchi identity for ${\cal R}_{ij}$ tells us that $g^{ik}\tilde{\nabla}_i\delta T_{kj}$ is a gradient, so we can introduce another potential $\Pi$ such that 
\begin{equation}
 \tilde{g}^{ik}\tilde{\nabla}_i\Pi_{kj}=\partial_j\Pi\;.
\end{equation}
The momentum-conservation equation (24) then reads
\begin{equation}
\Pi=-\delta p+2\dot{H}\,\delta N+\dot{U}+3HU\;.
\end{equation}
Using Eqs.~(21), (25), and (27), the field equation (22) now takes the form
\begin{eqnarray}
&&\dot{\Xi}^i{}_j+3H\Xi^i{}_j=\frac{1}{a^2}\left[\tilde{g}^{ik}\tilde{R}_{kj}-\frac{1}{4}\delta^i{}_j\tilde{g}^{kl}\tilde{R}_{kl}\right]\nonumber\\
&&~~~~~~+\tilde{g}^{ik}\Pi_{kj}+\frac{1}{2}\delta^i{}_j\Pi+O(a^{-4})\;,
\end{eqnarray}
where
\begin{equation}
\Xi^i{}_j\equiv \xi^i{}_j+\frac{1}{2}\delta^i{}_jU\;.
\end{equation}
As a check, we note that the field equation (20) may be put in  the form
\begin{equation}
\tilde{\nabla}_i\Big(\Xi^i{}_j-\delta^i{}_j\Xi^k{}_k\Big)+O(a^{-4})=0\;,
\end{equation}
which is automatically consistent with Eq.~(28)  as a consequence of Eq.~(26) and the Bianchi identity obeyed by $\tilde{R}_{ij}$. 

Up to this point we have kept open the possibility of anisotropic inertia terms of order $a^{-2}$ in order to emphasize the very great simplicities offered if anisotropic inertia can be neglected to this order.  If we now assume that $\Pi_{ij}=0$ to order $a^{-2}$, so that also $\Pi=0$ to this order, then Eq.~(28) reads  
\begin{equation}
\dot{\Xi}^i{}_j+3H\Xi^i{}_j=\frac{1}{a^2}\left[\tilde{g}^{ik}\tilde{R}_{kj}-\frac{1}{4}\delta^i{}_j\tilde{g}^{kl}\tilde{R}_{kl}\right]+O(a^{-4})\;.
\end{equation}
The general solution of Eq.~(31) is of the form
\begin{eqnarray}
&&\Xi^i{}_j({\bf x},t)=\left[{\cal G}^{ik}({\bf x}){\cal R}_{kj}({\bf x})-\frac{1}{4}\delta^i{}_j{\cal G}^{kl}({\bf x}){\cal R}_{kl}({\bf x})\right]\frac{1}{a^3(t)}\int_{T}^t a(t')\,dt'\nonumber\\&&~~~~~~~~~~+\frac{B^i{}_j({\bf x})}{a^3(t)}+O(a^{-4})\;,
\end{eqnarray}
where $T$ is any fixed time, ${\cal G}_{ij}({\bf x})$ is the value of $\tilde{g}_{ij}({\bf x},t)$ at that time,  ${\cal R}_{ij}({\bf x})$ is the Ricci tensor calculated from the 3-metric ${\cal G}_{ij}({\bf x})$; and $B^i{}_j({\bf x})$ is some function of ${\bf x}$ (and $T$), appearing in the solution of the homogeneous equation corresponding to Eq.~(31).  It is striking that once we assume an absence of vorticity and anisotropic inertia, all other terms that depend on the details of the energy-momentum tensor cancel in this solution. 

To solve for the metric, we need to complete our choice of gauge.  By using Eqs.~(18), (21), and (29), we have
\begin{equation}
\dot{\tilde{g}}_{ij}=2\tilde{g}_{ik}\Xi^k{}_j+\frac{2H^2}{\dot{H}}\tilde{g}_{ij}\Xi^k{}_k-\frac{H}{2a^2\dot{H}}\tilde{g}_{ij}\tilde{g}^{kl}\tilde{R}_{kl}+\tilde{g}_{ij}X+O(a^{-4})\;,
\end{equation}
where $X=O(1/a^2)$ depends on the matter perturbations
\begin{equation}
X\equiv -\frac{H}{\dot{H}}\Big(\delta T_{00}-2(3H^2+\dot{H})\delta N\Big)-U\left(1+\frac{3H^2}{\dot{H}}\right)\;.
\end{equation}
(There are other ways of writing Eq.~(33), with apparently simpler definitions of $X$, but as we will see the definition used here provides special advantages.)  Under a shift $t\rightarrow t+\epsilon({\bf x},t)$ in the time coordinate, with $\epsilon$ of order $1/a^2$ (and a corresponding transformation $x^i\rightarrow x^i+{\cal G}^{ij}\int dt\,a^{-2}\partial \epsilon/\partial x^j$ to keep $N_i=0$), the quantity $X$ undergoes the transformation
\begin{equation}
X\rightarrow X+2\frac{\partial}{\partial t}\Big(\epsilon H\Big)+\dots\;,
\end{equation}
so we can evidently choose $\epsilon$ to make $X=0$.  This choice is not unique; after choosing space and time coordinates so that $g_{i0}=0$  and $X=0$, we can preserve both conditions by making a further gauge transformation with 
\begin{equation}
t\rightarrow t+\tau({\bf x})/H(t)\;,~~~~~~~~x^i\rightarrow x^i+{\cal G}^{ij}({\bf x})\,\frac{\partial  \tau({\bf x})}{\partial x^j}\int             \frac{dt}{a^2(t)H(t)}\;,
\end{equation}
where $\tau$ is an arbitrary function only of ${\bf x}$.  This residual gauge freedom will be important in Sections V and VI.

In  a theory of multiple scalar fields $X$ is a linear combination of terms proportional to the scalar field perturbations and their time derivatives, so the choice $X=0$ is a generalization of the choice of gauge commonly used in studying single field inflation, in which the scalar field is unperturbed.  In general, the gauge choice $X=0$ provides the great advantage that we can solve Eq.~(33) for $\tilde{g}_{ij}$ without worrying  about the matter variables to which the  metric is coupled:
\begin{eqnarray}
&&\tilde{g}_{ij}({\bf x},t)={\cal G}_{ij}({\bf x})+2\left[{\cal R}_{ij}({\bf x})-\frac{1}{4}{\cal G}_{ij}({\bf x}){\cal G}^{kl}({\bf x}){\cal R}_{kl}({\bf x})\right]\int^t_T\frac{dt'}{a^3(t')}\int_{T}^{t'} a(t'')\,dt''\nonumber\\&&
~~~+\frac{1}{2}{\cal G}_{ij}({\bf x}){\cal G}^{kl}({\bf x}){\cal R}_{kl}({\bf x})\int_T^t\frac{H^2(t')\,dt'}{\dot{H}(t')a^3(t')}\int_{T}^{t'} a(t'')\,dt''\nonumber\\
&&~~~-\frac{1}{2}{\cal G}_{ij}({\bf x}){\cal G}^{kl}({\bf x}){\cal R}_{kl}({\bf x})\int_T^t \frac{H(t')\,dt'}{a^2(t')\,\dot{H}(t')}\nonumber\\&&
~~~+2\,{\cal G}_{ik}({\bf x})B^k{}_j({\bf x})\int_T^t \frac{dt'}{a^3(t')}+2\,{\cal G}_{ij}({\bf x})B^k{}_k({\bf x})\int_T^t \frac{H^2(t')\,dt'}{a^3(t')\,\dot{H}(t')}\nonumber\\&&~~~+O\Big(a^{-4}(t)\Big)\;,
\end{eqnarray}
where  $T$ is again any fixed time, and   ${\cal G}_{ij}({\bf x})$ and ${\cal R}_{ij}({\bf x})$ are the values of $\tilde{g}_{ij}$ and the associated Ricci tensor at that time.  This is the same solution found in [1] for  the special case of multiple scalar fields with an arbitrary potential.  In the language of Eq.~(1), this solution tells us that  
\begin{eqnarray}
&&f_1(t)=2\int^t_T\frac{dt'}{a^3(t')}\int_{T}^{t'} a(t'')\,dt''\\
&&f_2(t)=-\frac{1}{2}\int^t_T\frac{dt'}{a^3(t')}\int_{T}^{t'} a(t'')\,dt''\nonumber\\
&&~~~~~+\frac{1}{2}\int_T^t\frac{H^2(t')\,dt'}{\dot{H}(t')a^3(t')}\int_{T}^{t'} a(t'')\,dt''\nonumber\\
&&~~~~~-\frac{1}{2}\int_T^t \frac{H(t')\,dt'}{a^2(t')\,\dot{H}(t')}\;.
\end{eqnarray}
Of course, having chosen space coordinates so that $g_{i0}=0$, we have $f_3(t)=0$.  The calculation of $f_4(t)$ is taken up in Sections V and VI.

\begin{center} 
{\bf IV. Other Variables}
\end{center}

We have seen in Eq.~(34)  that the general solution of the dynamical equations for matter and metric variables that differs from  the ``trial configuration'' (for which $g _{ij}({\bf x},t)/a^2(t)={\cal G}_{ij}({\bf x})$, $g_{i0}=0$, $g_{00}=-1$, and all matter variables are unperturbed) by small terms, of order $a^{-2}$, has a spatial metric that approaches ${\cal G}_{ij}({\bf x})$ for large $a(t)$.  But in order to show that the trial configuration is truly attractive, we also need to consider the other metric component $g_{00}=-N^2$ (the coordinate system having been chosen so that $g_{i0}=0$), and the various matter variables.  We will do this in the next two sections for two cosmological models: single field inflation and local thermal equilibrium with no non-zero conserved quantities.  Both of these models have vanishing anisotropic inertia, and therefore have a spatial metric described by the results of the previous section.  From the perfect-fluid form of the energy-momentum tensor, we have in both cases
\begin{equation}
\delta T_{00}=\delta \rho+6H^2\delta N+O(a^{-4})
\end{equation}
so Eqs.~(21) and (29) give  
\begin{equation}
2\delta \rho=-a^{-2}\tilde{g}^{ij}\tilde{R}_{ij}+4H\Xi^k{}_k-6HU+O(a^{-4})\;,
\end{equation}
while Eqs.~(40) and (34) show that the gauge condition $X=0$ gives
\begin{equation}
0=-\frac{H}{\dot{H}}\Big(\delta\rho-2\dot{H}\delta N\Big)-U\left(1+\frac{3H^2}{\dot{H}}\right)
\end{equation}
Finally, for zero anisotropic inertia the equation (27) of momentum conservation gives
\begin{equation}
0=-\delta p+2\dot{H}\,\delta N+3HU+\dot{U}+O(a^{-4})\;.
\end{equation}
This gives three relations among the four quantities $\delta N$, $\delta\rho$, $\delta p$, and $U$, so with one additional relation we can calculate all four.

\begin{center}
{\bf V. Single Field Inflation}
\end{center}

In general, for a single scalar field $\bar{\varphi}(t)+\delta\varphi({\bf x},t)$, with a conventional kinematic term, the quantity $X$ defined by Eq.~(34) is given  by [1]: 
\begin{eqnarray}
&&X=-\dot{\bar{\varphi}}\delta \varphi+\frac{H}{\dot{H}}\Big(\ddot{\bar{\varphi}}\delta\varphi-\dot{\bar{\varphi}}\delta\dot{\varphi}\Big)+O(a^{-4})\nonumber\\&&~~~~=-\frac{1}{\sqrt{-2\dot{H}}}\left[-2\dot{H}\delta\varphi-
\frac{H}{\dot{H}}\Big(-\ddot{H}\delta\varphi+2\dot{H}\delta\dot{\varphi}\Big)\right]+O(a^{-4})\;.
\end{eqnarray}
The gauge choice that $X=0$ to order $a^{-2}$ tells us then that 
\begin{equation}
\delta\varphi({\bf x},t)=f({\bf x})\frac{\sqrt{-2\dot{H}(t)}}{H(t)}\;,
\end{equation}
where $f({\bf x})$ is some function only of ${\bf x}$.
Under a residual gauge transformation of the form (36) (which preserves the conditions $X=0$ and  $g_{i0}=0$) the scalar field perturbation is shifted by
$$\Delta\delta\varphi({\bf x},t)=-\dot{\bar{\varphi}}(t)\tau({\bf x})/H(t)=-\sqrt{-2\dot{H}(t)}\tau({\bf x})/H(t)\;,$$
so by choosing $\tau({\bf x})=f({\bf x})$, we can  make $\delta\varphi=0$, completing our choice of gauge.  The momentum potential defined by Eq.~(23) is here
\begin{equation}
U=\dot{\bar{\varphi}}\delta\varphi\;,
\end{equation}
so the gauge choice $\delta\varphi=0$ also entails $U=0$.  The density and pressure perturbations here are
\begin{equation}
\delta \rho=\delta p = \frac{1}{2}\delta g_{00}\dot{\bar{\varphi}}^2=2\dot{H}\delta N\;,
\end{equation}
so Eqs.~(42) and (43) are satisfied, and the only variable that needs to be examined to check the attractive nature of the adiabatic solution is $\delta g_{00}=-2\delta N$.  It is given by Eqs.~(47) and (41) as  
\begin{eqnarray}
&&\delta N=\frac{1}{4\dot{H}}\left[-a^{-2}\tilde{g}^{ij}\tilde{R}_{ij}+4H\Xi^k{}_k\right]+O(a^{-4}) \nonumber\\&&~~~~~~
=\frac{{\cal G}^{ij}{\cal R}_{ij}}{4\dot{H}}\left[-\frac{1}{a^2}+\frac{H}{a^3}\int_T^t a(t')\;dt'\right]+O(a^{-3})\;.
\end{eqnarray}
This is $O(a^{-2})$, and hence vanishes for large $a$, so the adiabatic solution is indeed an attractor in the case of single field inflation.  In the language of Eq.~(1), Eq.~(48) gives
\begin{equation}
f_4(t)=-\frac{1}{2\dot{H}(t)}\left(-\frac{1}{a^2(t)}+\frac{H(t)}{a^3(t)}\int_T^t a(t')\;dt'\right)\;,
\end{equation}
In the gauge adopted here there are no matter perturbations during single field inflation, and so nothing else to check.

\begin{center}
{\bf VI. Local Thermal Equilibrium}
\end{center}
 
In thermal equilibrium with no non-zero conserved quantum numbers the pressure and energy density in any gauge are functions only of the temperature, so
\begin{equation}
\delta p=(\dot{\bar{p}}/{\dot{\bar{\rho}}})\,\delta\rho=\left(-1-\frac{\ddot{H}}{3H\dot{H}}\right)\,\delta\rho\;.
\end{equation}
Using this and Eqs.~(41) and (42)  in Eq.~(43) gives a differential equation for the momentum potential $U$:

\begin{eqnarray}
&&\dot{U}+U\,\left(\frac{\dot{H}}{H}-\frac{\ddot{H}}{\dot{H}}\right)=
-\left(2+\frac{\ddot{H}}{3H\dot{H}}\right)\left(-\frac{1}{2a^2}\tilde{g}^{ij}\tilde{R}_{ij}+2H\Xi^i{}_i\right)
\nonumber\\&& ~~~ = -\left(1+\frac{\ddot{H}}{6H\dot{H}}\right)\left(-\frac{1}{a^2}+\frac{H}{a^3}\int_T^t a(t')\,dt'\right){\cal G}^{ij}{\cal R}_{ij}+O(a^{-3})\;,
\end{eqnarray}
where $T$ may be taken as any time during the period of thermal equilibrium, most conveniently at its beginning, and ${\cal G}_{ij}$ and ${\cal R}_{ij}$ are the reduced metric $g_{ij}/a^2$ and associated Ricci tensor at that time.  The solution is
\begin{eqnarray}
&&U({\bf x},t)=f({\bf x})\frac{\dot{H}(t)}{H(t)} \nonumber\\&&
~~~-{\cal G}^{ij}({\bf x}){\cal R}_{ij}({\bf x}) \frac{\dot{H}(t)}{H(t)} \int_T^t dt'\; \left(\frac{H(t')}{\dot{H}(t')}\right)\left(1+\frac{\ddot{H}(t')}{6\,H(t')\,\dot{H}(t')}\right)\nonumber\\&&~~~~~~\times
\left(-
\frac{1}{a^2(t')}+\frac{H(t')}{a^3(t')}\int_T^{t'}a(t'')\,dt''\right)+O(a^{-3})\;,
\end{eqnarray}
where $f({\bf x})$ is an arbitrary function of position.
At first sight, this does not appear very attractive, in either the mathematical or the colloquial sense.  The first term, representing a solution of the homogeneous equation corresponding to (51), is of zeroth order in $1/a$, and so does not become small for large $a$.  But this term is a gauge artifact.  From the definition (23) of the momentum potential, we can easily see that under the residual gauge transformation (36) (which preserves the conditions $X=0$ and $g_{i0}=0$), $U$ transforms as
\begin{equation}
U({\bf x},t)\rightarrow U({\bf x},t)+\left(\frac{2\dot{H}(t)}{H(t)}\right)\tau({\bf x})
\end{equation}
so by choosing the arbitrary function $\tau({\bf x})$ to have the value $-f({\bf x})/2$, we can cancel the first term in Eq.~(52).  The remainder is of order $a^{-2}$, and hence vanishes for large $a$.  

The remaining perturbations $\delta g_{00}=-2\delta N$ and $\delta\rho$ are algebraically related to $U$ by Eqs.~(41) and (42), which give 
\begin{equation}
\delta\rho({\bf x},t)=-3H(t)U({\bf x},t)+\frac{1}{2}{\cal G}^{ij}({\bf x}){\cal R}_{ij}({\bf x})\left(-\frac{1}{a^2(t)}+\frac{H(t)}{a^3(t)}\int_T^t dt'\;a(t')\right)
\end{equation}
and
\begin{equation}
2\dot{H}(t)\delta N({\bf x},t)=\left(\frac{\dot{H}(t)}{H(t)}\right)U({\bf x},t)+\frac{1}{2}{\cal G}^{ij}({\bf x}){\cal R}_{ij}({\bf x})\left(-\frac{1}{a^2(t)}+\frac{H(t)}{a^3(t)}\int_T^t dt'\;a(t')\right)\;.
\end{equation}
Both are of order $a^{-2}$, so in this case, too, the adiabatic solution is completely attractive for large $a$.
In the language of Eq.~(1), Eqs.~(55) and (52) give:
\begin{eqnarray}
&&f_4(t)=\frac{\dot{H}(t)}{H^2(t)}\int_T^t \left(\frac{H(t')}{\dot{H}(t')}+\frac{\ddot{H}(t')}{6\dot{H}^2(t')}\right)\left(-\frac{1}{a^2(t')}+\frac{H(t')}{a^3(t')}\int_T^{t'}a(t'')\,dt''\right)\;dt'\nonumber\\&&~~~~
-\frac{1}{2\dot{H}(t)}\left(-\frac{1}{a^2(t)}+\frac{H(t)}{a^3(t)}\int_T^{t}a(t')\,dt'\right)\;.
\end{eqnarray}

\vspace{12pt}

I am grateful for helpful conversations with  Raphael Flauger and Eiichiro Komatsu.  This material is based upon work supported by the National Science Foundation under Grant No. PHY-0455649 and with support from The Robert A. Welch Foundation, Grant No. F-0014

\begin{center}
{\bf References}
\end{center}

\begin{enumerate}

\item S. Weinberg, arXiv:0808.2909.

\item Some aspects of these results have been obtained by  D. H. Lyth, K. A. Malik, and M. Sasaki, J. Cosm. Astropart. Phys. {\bf 05}, 004 (2005).
The $O(a^{-2})$ and $O(a^{-3})$ terms in the metric have also been found by
Y. Tanaka and M. Sasaki, Prog. Theor. Phys. {\bf 181}, 455 (2007) in  the special case of single-field inflation.  Their solution is different from that presented in Section III, presumably because they use a different gauge.

\item For tensor modes, see S. Weinberg, Phys. Rev. D {\bf 69}, 023503 (2004).  For scalar modes, see Eq. (6.1.62) of S. Weinberg, {\em Cosmology} (Oxford University Press, 2008).

\item R. S. Arnowitt, S. Deser, and C. W. Misner, in {\em Gravitation: An Introduction to Current Research}, ed. L. Witten (Wiley, New York, 1962): 227, now also  available as gr-qc/0405109.

\end{enumerate}

\end{document}